\documentclass[
]{ceurart}

\usepackage{listings}
\begin{document}

\copyrightyear{2025}
\copyrightclause{Copyright for this paper by its authors.
  Use permitted under Creative Commons License Attribution 4.0
  International (CC BY 4.0).}

\conference{The Third Workshop on Building an Inclusive and Accessible Metaverse for All, 26 April-1 May 2025, Yokohama, Japan}

\title{Conducting VR User Studies with People with Vision/Hearing Impairments: Challenges and Mitigation Strategies}


\author{Wenge Xu}[%
orcid=0000-0001-7227-7437,
email=wenge.xu@bcu.ac.uk,
]
\cormark[1]
\address{HCI Research Centre, College of Computing, Birmingham City University, Birmingham, United Kingdom}

\author{Craig Anderton}[
orcid=0000-0002-7172-217X,
email=craig.anderton@bcu.ac.uk,
]

\author{Kurtis Weir}[
orcid=0000-0002-7094-4600,
email=Kurtis.Weir@bcu.ac.uk,
]

\author{Arthur Theil}[
orcid=0000-0002-8900-8740,
email=arthur.theil@bcu.ac.uk,
]


\cortext[1]{Corresponding author.}

\begin{abstract}
There is a lack of virtual reality (VR) user studies that have been conducted involving people with vision/hearing impairments. This is due to the difficulty of recruiting participants and the accessibility barriers of VR devices. Based on the authors' experience conducting VR user studies with participants with vision/hearing impairments, this position paper identifies 5 key challenges (1. Recruitment, 2. Language Familiarity, 3. Technology Limitations and Barriers, 4. Access to Audio Cue, and 5. Travelling to the Experiment Location) and proposes strategic approaches to mitigate these challenges. In addition, we also presented three key considerations regarding understanding participants' lived experiences that could help the user study become accessible.
\end{abstract}


\maketitle

\section{Introduction}
The lack of diversity in participants in the human-computer interaction (HCI) user study field is a known issue, especially for people with disabilities \cite{diversity_in_study_hci,hci_sample_detail}. Accessibility research is a main area where participants with disabilities participate \cite{mapping_20_accessibility}. According to a review of accessibility papers published at CHI (2010-2019) and ASSETS (1994-2019) by Mack et al. \cite{Mack_accessibility_review}, among 477 accessibility papers with user studies, about 90$\%$ of the accessibility research included participants with disabilities and/or older adults. However, the percentage of such accessibility studies published each year is still relatively low. For instance, it only makes up 7.8$\%$ (55/702) of all published papers at CHI in 2019.

Similar to the broader HCI user studies, there is a lack of (1) VR accessibility studies and (2) participation of participants with disabilities in VR-based user studies \cite{9646525}. Several attempts have been made to address this issue in VR-based user studies. Dr John Quarles delivered a keynote session \cite{9089564} at the 2020 IEEE Virtual Reality Conference (IEEE VR'20) to explain why the current lack of diversity in VR is a major threat to the generalisability of VR research and could be a significant barrier to the success of the VR industry. Following the keynote, they further produced a call for action to increase the diversity of study participants for VR studies \cite{9646525}. On the other hand, there are efforts from the extended reality conference community, where Diversity and Inclusion of Diverse User Populations have been placed as a criterion (e.g., IEEE ISMAR). Nevertheless, there is still limited research conducted on people with disabilities. 

The lack of VR accessibility research and diversity in VR participants is not just due to the difficulty in recruiting participants with disabilities \cite{10.1145/1978942.1979268}; the accessibility barriers of devices have worsened it \cite{Creed17102024,10.1007/s10055-023-00850-8}. The goal of this position paper is to explore the most prevalent challenges in conducting VR-based user studies with participants with vision/hearing impairments and propose strategic approaches to mitigate these challenges. Reflections and recommendations from this work should guide future research on how to include participants with disabilities and make their user studies accessible. 

\section{Key Challenges and Their Mitigation Strategies}
We consider the following challenges that require attention if a researcher wants to conduct VR user studies with participants with vision/hearing impairments. Each challenge comes with mitigation strategies based on our experience.

\subsection{Recruitment}
As recruitment is often the most common challenge in conducting VR user studies with people with disabilities, the following strategies are recommended:

\begin{itemize}
  \item \textbf{Charities and other Non-profit Organisations}. Working with local, regional, and national charities and other non-profit organisations would be a good option for participant recruitment. These organisations may already have channels/communities to post participant recruitment information. However, contacting them earlier and building relationships with the charities would be needed; charities may request different lead times to publish the recruitment post to their channel/community.
  
  \item \textbf{Online Forum/Community}. Charities and other non-profit organisations have more in-person connections, online forums, and communities where people with disabilities can engage. Engaging with online forums used by those communities (e.g., vision loss communities, Deaf communities, hearing-aid communities) would provide you extra opportunities to start the conversation and recruit them for the user study. 

  \item \textbf{Timely Connections}. We suggest that researchers get involved with research candidates much earlier to build connections, in turn potentially increasing the likelihood of further recruitment through snowball sampling. Participants should be encouraged to spread recruitment information to potential candidates if they agree on the value of the study for their community.

  \item \textbf{Lived Experience within the Team}. When formulating the research team, having members that can represent the target audience from the very start of a study can accelerate collaborations between other organisations and persons, while also providing lived perspectives from the earliest stages that will feed directly into the study design.
  
  \item \textbf{Build Your Reputation}. Regardless of the above methods, the researcher should have the mindset that it is not intended just to get people with disabilities for the user study only; it is required to get to know them, stay connected, and volunteer for their communities and charities would help them and the research team itself in the long term. Be part of their community/forum, learn their rules and interests, and become interested in what they do/say. Share the perspective, the story, and why the researcher is interested.
\end{itemize}

\subsection{Communication: Written/Spoken Language Familiarity}\label{communication_Deaf}
This challenge could induce barriers such as (1) reading written documents (information sheets, consent forms, and questionnaires), (2) fill out the consent form and text-based questionnaires, and (3) communicate with the experimenter (general experiment procedure, interview, emails). 

The written/spoken language may not be the first language for sign language users. Unclear or unintelligible written content for sign language users could result in biased and unethical responses \cite{Deaf_ethics}. Most critically, many sign language users may not be able to read the written content, further creating barriers for them to attend the study and resulting in them being excluded from the research due to the lack of accessible informed consent and research materials. It should be noted that written content could still be challenging for sign language users who can read written text as a secondary language \cite{Written_language_proficiency}, as the sign language is different from the written language \footnote{\url{https://www.signsolutions.uk.com/why-do-some-deaf-people-struggle-with-written-text/}}. It may result in an increase in the duration of the experiment due to potential barriers in filling out the questionnaire, reading the information sheet, and signing the consent form. The following are our recommendations:

\begin{itemize}
  \item \textbf{Employing a Sign Language Interpreter}. A sign language interpreter would be necessary if the experimenter does not know sign language. This would help with communication of all kinds. The intensity of the communication within an experiment needs to be carefully considered, as sign language interpretation is a fatiguing job. Suppose the sessions require a lot of real-time translation (e.g., a brief of the study, instructions, questionnaire, and interview). In that case, there is a need to have multiple sign language interpreters to make the communication smooth or have pre-recorded study briefs and questionnaires to minimise the workload of the sign language interpreter. Employing sign language interpreters may not address the challenge entirely, especially if they are unfamiliar with the field; hence, contacting them beforehand with relevant materials or arranging a quick call with the interpreter would be needed.

    There is ongoing research and development on developing automated sign language systems (translating text/voice to sign language) \cite{sign_language_avatar1,10.1145/3527188.3561923} and sign language recognition systems (translating sign languages to text) \cite{Sign_language_recognition}. However, the quality and efficiency of employing these solutions for a real-time user study remain unclear due to a lack of investigations in the field to show the feasibility of these tools in a user study, but also known challenges identified in \cite{Sign_language_recognition}, e.g., recognition of continuous sign language speech, limited vocabularies, and understanding the sign language communications. We would hesitate to recommend this in this current position paper.

  \item \textbf{Translating Written Document in Sign Languages}. Materials should be translated into sign language videos (e.g., questionnaires \cite{deaf_questionnaire}, procedure, study description, consent form, information sheet). This can save the interpreter's energy, but most importantly, this may be helpful for short-duration studies that may need two interpreters, as there is a shortage of sign language interpreters worldwide. 

  \item \textbf{Make Yourself Visible}. Regardless if the participant is capable of lip reading, always ensure the face is facing towards the participant. In addition, non-verbal communication such as facial expressions, eye contact, and body language can both help with communication\footnote{\url{https://www.hearinglink.org/living/lipreading-communicating/non-verbal-communication/}}, therefore, ensuring the researcher's lip, face, and body are visible to the participants at all times during the communication with the participants is essential.
\end{itemize}

\subsection{Communication: Technology Limitations and Barriers}
Dealing with written documents such as information sheets, consent forms, and text-based questionnaires is also a challenge for people with vision impairment. This is mainly due to the limitation of the technologies and access barriers of the technologies. 

\textbf{Technology Limitations}. People with vision impairment use screen reader in their daily lives for "reading" the content on the devices; despite advancements in screen reader software, challenges persist in navigating complex web-based content \cite{screen_reader_web}, charts \cite{screen_reader_charts}, images \cite{screen_reader_image}, and online applications \cite{screen_reader_online}. In addition to reading content, user studies involve consent form, demographic questionnaire, and qualitative questions that require participants to respond via text. A typical technology people with vision impairment rely on is voice input technology. However, there are issues with such technology (e.g., accuracy, data privacy, online connection) \cite{10.1145/3529190.3529197}, which may affect user studies. Confirming whether the content has been correctly input from participants with vision impairment may also be time-consuming as they would rely on a screen reader. However, not checking the inputted content may result in researchers experiencing difficulty in understanding the feedback from participants due to the potential transcription issues introduced by speech recognition errors.

\textbf{Technology Barriers}. Older people with vision impairment often do not engage with technological solutions such as the above-mentioned screen readers and voice input, causing them to share the same issues listed in Section \ref{communication_Deaf}. They may (1) decide not to read documents themselves due to the difficulty of seeing the documents, (2) not be capable of filling out the forms themselves, and (3) communicate with the researcher via any text-based communication media (email, message). 

For both challenges, we recommend the following strategy:
\begin{itemize}
    \item \textbf{Being Flexible/Providing Options}. Screen reader and voice input technologies can still be provided depending on participants' preferences and with the help of the researcher. It would be suitable for the experimenter also to be flexible and provide options such as signing the consent form in person, leaving more time for the participant to deal with the document, printing documents with large font size, reading the written documents out for the participants. 
    
    As for filling out the consent form and questionnaires, discuss with the participant beforehand to explore options such as (1) through a third party (charity, friend, family member) to fill for them, or (2) let the researcher to fill on behalf of the participant.

    Providing people with multiple options and being flexible would help to both make the user study more accessible, along with mitigating potential issues with assistive technology devices by minimising the potential of a single point of failure. 
\end{itemize}

\subsection{Access to Audio Cue}
We rely on audio cues from the surroundings to understand the real world while immersed in the virtual world. However, audio cues are not accessible to people with severe/profound hearing loss levels. We recommended the following mitigation strategies:

\begin{itemize}
    \item \textbf{Clarify Things Beforehand}. Before putting on the head-mounted display (HMD), instructions about the user study (task, control, duration, etc.) should be clarified to avoid communication barriers caused by the lack of access to audio cues while participants wear the HMD. 

    \item \textbf{Video See-Through}. If there is a compulsory requirement to explain things while the participant is wearing the HMD, the researchers should employ an HMD with good video see-through capabilities \cite{5643530,10.1561/1100000049}. A good video see-through capability would allow participants with severe/profound hearing loss levels to engage in a potential conversation with the experimenter, helping to navigate who is speaking and providing visual access to (1) faces for lip reading and facial expression, and (2) interpreters providing sign language.  
\end{itemize}

\subsection{Travelling to the Experiment Location}
Many VR studies require participants to attend in person. There are several barriers for people with vision impairment when travelling via public transportation \cite{barriers_vision_travel}. The best case would be inviting participants living in the local area, who may experience fewer barriers in familiar environments. However, getting enough sample size for the participants is typically impossible and insufficient. Strategies to mitigate potential travel barriers include:

\begin{itemize}
    \item  \textbf{Remote VR User Study}. Remote user studies could be offered if the participants can conduct the study by themselves safely. Remote user studies in VR have gained significant attention since the COVID-19 pandemic, and several papers have been published sharing the best practices and identifying potential challenges \cite{10.1145/3568444.3568462, paneva2024exploringvulnerabilitiesremotevr}. 

    \item \textbf{Accompanying the Participant}. Escorting the participants to the dedicated user study location would be ideal. Researchers should undergo sighted guide training to become familiar with the techniques and processes of sighted guides\footnote{\url{https://www.rnib.org.uk/living-with-sight-loss/supporting-others/guiding-a-blind-or-partially-sighted-person/}}. 

    \item \textbf{Renting/Negotiating a Local Space}. Local charities/communities may have a space that can be occupied. Providing the space is sufficient for the requirements of the study, the researcher can discuss with the local charity/community to explore solutions, e.g., booking a room for a certain period.
    
    \item \textbf{Active Travel of the Experimenter}. When the above strategies are infeasible, the experimenter may have to travel to the participant's residence to conduct in-home research. However, this could increase the risks for the experimenter; care should be taken for the experimenter themselves \cite{in_home_research}; see these slides for important considerations of conducting in-home research towards university ethics community\footnote{\url{https://research.uga.edu/docs/units/hso/Important_Considerations_in_Conducting_In_home_Research___8_19_22.pdf}}. 
\end{itemize}

\section{Understanding Participants' Lived Experience: Key Considerations}
In addition to the above-mentioned challenges, we present the following three good practices that should be considered when planning to conduct a VR-based user study with participants with vision/hearing impairments:

\begin{itemize}
    \item \textbf{Uncomfortable Interactions}. Check whether the study design would involve specific interactions that may not be comfortable for participants based on their own lived experiences, such as repetitive movements that may resemble self-stimulation behaviours. Refer to literature, consultation, or a pilot study to better understand the emotional complexities tasks may evoke for participants. Consider adapting or minimising such tasks if possible.
    
    \item \textbf{Clear Instructions}. Due to the potential unfamiliarity with VR technology and user studies procedures, participants with vision/hearing impairments may be unaware of the demands of the experiment (procedure, task, interaction). Researchers should be very clear about the procedure and expected interactions, how long participants are required to wear the HMD in a single period, and how long participants can rest between conditions, including allowing for additional resting periods or overall testing duration according to individual needs.

    \item \textbf{Translation of Designs}. Linking to the previous points, many commercial VR setups take design principles from video game systems, which can contribute towards a participants' unfamiliarity to not only physical interactions (e.g., gamepad controllers), but also gamified concepts. Participants with vision/hearing impairments, especially ones that are older generation, have shown to struggle when trying to adapt to gamified systems that they are not familiar with. It is crucial researchers consider the influences of expected VR design and how to translate these towards a wider audience that can have an unfamiliarity with gaming devices.
    
\end{itemize}

\section{Conclusion}
Virtual reality (VR) has great potential in learning, working, leisure, and entertainment. However, there is a significant lack of diversity among VR user study participants; only limited studies have been conducted involving participants with disabilities. In this position paper, based on our experience conducting VR user studies with participants with vision/hearing impairments, we identified 5 challenges and proposed mitigation strategies for each. Together, we also presented 3 key considerations regarding understanding participants' lived experiences that could help the study become accessible. As the technologies (VR, screen reader, voice input) continue to evolve, ongoing reviews to identify and mitigate challenges will be crucial in ensuring the accessibility of conducting VR user research.

\begin{acknowledgments}
This work is supported by Royal Society Research Grant ($RG\backslash R1\backslash 241114$).
\end{acknowledgments}

\bibliography{chiea}

\end{document}